\documentstyle[preprint,aps]{revtex}
\def\L{\Lambda}
\def\bea{\begin{equation}}
\def\ena{\end{equation}}
\def\bey{\begin{eqnarray}}
\def\beyn{\begin{eqnarray*}}
\def\eny{\end{eqnarray}}
\def\enyn{\end{eqnarray*}}
\def\ds#1{\ooalign{$\hfil/\hfil$\crcr$#1$}}
\begin{document}
\begin{titlepage}
\begin{flushright}
IFUM 585/FT\\
revised version\\
January 1998\\
\end{flushright}
\vspace{1.5cm}
\begin{center}
{\bf \large HARD--SOFT RENORMALIZATION AND THE EXACT RENORMALIZATION GROUP}
\footnote{Work supported in part by M.U.R.S.T.}\\
\vspace{1 cm}
{ M. PERNICI} \\ 
\vspace{2mm}
{\em INFN, Sezione di Milano, Via Celoria 16, I-20133 Milano, Italy}\\
\vspace{0.6 cm}
{ M. RACITI and F. RIVA}\\
\vspace{2mm}
{\em Dipartimento di Fisica dell'Universit\`a di Milano, I-20133 Milano,
Italy}\\
{\em INFN, Sezione di Milano, Via Celoria 16, I-20133 Milano, Italy}\\
\vspace{2cm}
\bf{ABSTRACT}
\end{center}
\begin{quote}
\tightenlines
The Wilsonian exact renormalization group gives a natural framework in which
ultraviolet and infrared divergences can be treated separately.
In massless QED we introduce, as the only mass
parameter, a renormalization scale $\L_R > 0$.
We prove, using the flow equation technique, that infrared
convergence is a necessary consequence
of any zero-momentum renormalization condition at $\L_R$ compatible with
the effective Ward identities and axial symmetry.
The same formalism is applied to renormalize gauge-invariant
composite operators and to prove their infrared finiteness;
in particular we consider the case of the axial current operator and 
its anomaly.

\end{quote}
\end{titlepage}
\tightenlines
\section*{Introduction}
The quantization of massless theories requires the introduction of a
mass scale which, in perturbation theory, can appear for instance in the renormalization conditions
as a non-zero momentum subtraction point \cite{Rm}.
In these theories one has to deal both with infrared and ultraviolet 
divergences and it is convenient to choose a renormalization scheme
allowing a separate treatment of these divergences.
The frequency-splitting Wilsonian approach to renormalization
\cite{Wil} yields an elegant way of making this separation.

In  \cite{LM} this splitting is realized in massless $g\phi^4$ 
replacing the scalar field with hard and soft fields, which roughly 
propagate respectively above and below the renormalization scale
$\L_R$; it is shown that in the hard-soft theory 
the renormalization conditions can be taken at zero momentum, allowing
a simpler BPHZ renormalization than using a non-zero momentum
subtraction point.

Making a momentum decomposition with a continuous scale $\L$,
the Wilsonian Green functions  
satisfy an exact renormalization flow equation \cite{Wil}, using which
Polchinski \cite{Polchi} gave  a simple proof of renormalizability 
in massive  $g\phi^4$ 
for Green functions with momenta smaller than $\L_R$.
In \cite{KKS} the same equation has a complementary meaning; it
describes indeed the variation of the Green functions computed with 
an infrared cut-off $\L$, as $\L$ varies from the ultraviolet
cut-off $\L_0$ to  $0$; technical semplifications of the  proof in
\cite{Polchi}  were moreover obtained.
This proof has been extended to the massless case in \cite{BDM,KK}.
In \cite{Morris} it was noticed that the \cite{Polchi} and \cite{KKS}
approaches can be dealt together using a hard-soft field
decomposition. 

It is clear that the approach \cite{LM} is related to the one in
 \cite{Polchi,KKS,BDM,KK,Morris}; to our knowledge this
relation has not been discussed in the literature. 
In both cases the hard-soft decomposition at scale $\L_R$  allows the 
choice of zero-momentum renormalization conditions, which we will call
in the following hard-soft (HS) renormalization schemes.

The main differences between the two approaches are:
i) the hard-soft field decomposition  depends on the cut-off function chosen; 
in \cite{LM}  the cut-off function is ${\L_R^2\over p^2 +\L_R^2}$, so
that the soft theory is super-renormalizable; in  \cite{Polchi} and
\cite{KKS} a smooth compact-support cut-off function is chosen, and the
corresponding soft theory is ultraviolet-finite. 
ii) from a technical point of view, in \cite{LM} $\L$ is kept fixed at
$\L_R$ and the proof of renormalizability is made using the standard
BPHZ results, while the analogous proof in \cite{Polchi,KKS}
is made using the flow equation for varying $\L$.

Apart from renormalizability, other issues have been studied
 independently in these two approaches.

 In  \cite{LM} the $\L_R$-dependence of the Green functions is
controlled by a renormalization group equation, proved using the
Quantum Action Principle; in  the Polchinski approach 
it is controlled similarly by the renormalization flow equation \cite{HL}.

The application of these methods to the case of gauge theories might
seem difficult, since the hard-soft decomposition is not gauge invariant;
however, in the approach of \cite{LM}, it has been shown in \cite{Val} that,
 after this splitting, the original BRS invariance  is replaced by a 
generalized non-local BRS symmetry in the hard-soft theory. 
Analogously, Polchinski's approach has been extended to the
case of gauge theories in \cite{Becchi2,bon1,bon2,bon3,KKQED1,KKQED2}. 
In \cite{Becchi2} it has been shown that in Yang--Mills theory there are 
effective Ward identities for the Wilsonian action at scale $\Lambda$. The
corresponding Slavnov--Taylor identities on the effective $1$-PI
functional generator have been written down in \cite{El,DM} .
We will show that these identities follow indeed from  generalized non-local
BRS symmetry in the hard-soft theory; in the particular case in which
the cut-off function is  ${\L_R^2\over p^2 +\L_R^2}$ and the Feynman
gauge is chosen, this is the non-local BRS symmetry discussed in \cite{Val}. 

In \cite{Becchi2} the effective Ward identities have been proven in a
HS scheme together with renormalizability 
without dealing with the infrared problem.
This problem has been addressed using the flow equation
in \cite{bon3,KKQED2}, where  the renormalization conditions have
been chosen more conventionally at $\L=0$ and at a non-zero momentum 
subtraction point; in this scheme
the infrared and ultraviolet problems must be solved together, loop by loop.

Although this solution of the infrared problem guarantees, using
the flow equation, its solution in the HS
schemes, it would be simpler and more natural to give a direct proof in the
latter schemes, relying only on the effective Ward identities of 
\cite{Becchi2}: infrared finiteness and the usual Ward identities
should 
follow simply from this requirement.
Another motivation for a direct solution of the infrared problem in
the HS schemes, already advocated in \cite{LM}, is that it is
technically easier to make subtractions at zero-momentum rather than at a
non-zero subtraction point; this fact is particularly
relevant when the Ward identities cannot be trivially
maintained with a suitable choice of regularization.

In this perspective it is interesting  to investigate further the HS schemes; 
to make these schemes useful tools, one should be able to reproduce the
main results in QFT without making any reference to other more
traditional renormalization schemes;
first of all, it is necessary to prove the infrared finiteness.

In this paper we consider massless QED as a simple model to
implement this program.
We outline a proof of infrared finiteness in massless QED in a HS
scheme with smooth cut-off, using the flow equation technique. 
 There are only two relevant couplings
which are not fixed by symmetry requirements at $\L_R$; for any value
of them the theory is infrared finite.
In fact the proof is in some way simpler than in massless $g\phi^4$,
where the relevant coupling of dimension two must be
fine-tuned  at $\L_R$ to ensure infrared finiteness. 

We discuss gauge-invariant composite operators using
a variant, in the HS scheme, of the Zimmermann definition of the 
normal products.
In particular we define the gauge-invariant axial current operator and
we show how the axial anomaly appears in this context.

In the first section we discuss the hard-soft decomposition 
in gauge theories, showing that the gauge (or BRS) symmetry is
replaced by a non-local symmetry between the hard and soft fields,
along the lines of\cite{Val}, leading to the
effective Ward identities found subsequently in \cite{Becchi2,El,DM} on
the Wilsonian action.
In the second section we make one-loop computations in massless QED
using a HS renormalization scheme.
In the third section we use the effective Ward identities in
massless QED to prove its infrared convergence in a HS scheme 
using the exact renormalization flow equation.
We discuss the effective axial Ward identity in terms of
Zimmermann-like normal operators.
The last section contains remarks and the conclusion.

\section{Hard-soft renormalization in gauge theories}
\subsection{Scale decomposition}
In this subsection we review the hard-soft field formalism along the lines
of \cite{LM},\cite{ZJ} and \cite{Morris}; our presentation holds also in the
case of momentum cut-off functions with compact support.

Consider a massless field theory in Euclidean four-dimensional space, 
with classical action 
$$ 
S_{cl}(\Phi) = {1\over2}\Phi D^{-1}\Phi + S^I_{cl}(\Phi)
$$
where we use a compact notation in which $\Phi\equiv\{\Phi_i\}$ is a vector corresponding to a 
collection of fields and the index $i$ includes the space-time variables;
 $D \equiv \{D_{ij}\}$
 is the propagator matrix. $ D_{ij}= (-1)^{\delta_j}D_{ji}$ where
$\delta_j = 0 $ (or $1$) for an (anti)commuting field $\Phi_j$. 
 In all inner products like $\Phi D^{-1}\Phi\equiv \Phi_i D^{-1}_{ij} \Phi_j$
the inner product symbol is understood.

The bare action has the form:
\bea\label{act}
S_{\L_0}(\Phi) = {1\over2}\Phi D_{0\L_0}^{-1} \Phi + S^I_{\L_0}(\Phi)
\ena
where $D_{0\L_0}=DK_{\L_0}$; $\L_0$  is the ultraviolet cut-off.

The cut-off function $K_{\Lambda}(p)= K\left({p^2\over \L^2}\right)$ 
can be defined
in various ways; for some purposes it is convenient to define it on
a compact support (this is the choice made by Polchinski in his
proof of renormalizability of $\phi^4$ using the exact renormalization
flow \cite{Polchi})  or it can be defined analytic, as long as it goes to zero
at least as $\Lambda^4/p^4$ for large momentum and $K(0) = 1$; we will
use such a cut-off in the next section.
 
 At tree level 
$S^I_{\L_0}(\Phi)= S^I_{cl}(\Phi)$.
At quantum level the theory is characterized by a renormalization mass
scale $\L_R$.
The usual functional generator is
\bea\label{ZZ}
Z_{0\L_0}[J] = e^{W_{0\L_0}[J]} =  
\int{\cal D}\Phi e^{- S_{\L_0}(\Phi) + J \Phi}
\ena
The corresponding usual $1$-PI functional generator is called 
$\Gamma_{0\L_0}[\Phi]$.

Define now fields $\Phi_S $ and $\Phi_H $ on the supports of
\bey
K_{\L_R} \equiv K_S \qquad 
K_{\L_0}-K_{\L_R}\equiv K_H
\eny
respectively, with propagators
\bey\label{prop}
D_{0\L_R} = K_{\L_R}D \equiv D_S\qquad  D_{\L_R \L_0}
=(K_{\L_0}-K_{\L_R})D \equiv D_H
\eny

 Using gaussian integration one can decompose the usual functional
integral in the following way:
\bea\label{Z}
Z_{0\L_0}[J] =
N \int {\cal D}\Phi_S {\cal D}\Phi_H e^{-S_{\L_0}(\Phi_S,\Phi_H)+J(\Phi_S+\Phi_H)}
\ena
where $N$ is a normalization constant and
\bea\label{action}
S_{\L_0}(\Phi_S,\Phi_H) =
{1\over2} \Phi_SD_S^{-1}\Phi_S +{1\over2} \Phi_HD_H^{-1}\Phi_H +
S^I_{\L_0}\left(\Phi_S +\Phi_H  \right)
\ena
The Wilsonian path-integral, in which only the high modes $\Phi_H$
are integrated out, is 
\bey
Z_{\Lambda_R \Lambda_0}[J,\Phi_S]= \exp{W_{\L_R\L_0}}[J,\Phi_S]
= \int {\cal D} \Phi_H e^{-S_{\L_0}(\Phi_S,\Phi_H)+J\Phi_H}
\eny

Making a Legendre transformation from $W_{\Lambda_R \Lambda_0}$
to $\Gamma_{\Lambda_R \Lambda_0}$ one arrives at an expression
for the effective action of the form
\bea \label{Gint}
\Gamma_{\L_R \L_0}[\Phi_S,\Phi_H^c]
= \frac {1}{2} \Phi_S D_S^{-1} \Phi_S +\frac {1}{2} \Phi_H^c
D_H^{-1}\Phi_H^c
+\bar\Gamma^{int}_{\L_R \L_0} \left[\Phi_S,\Phi_H^c \right]
\ena
where
\bea\label{one}
\bar\Gamma^{int}_{\L_R \L_0} \left[\Phi_S, \Phi_H^c \right] =
\Gamma^{int}_{\L_R \L_0} \left[ \Phi_S + \Phi_H^c\right]
\ena
In fact (for $p$ in the intersection of the supports of
$ K_H $ and $K_S$):

\beyn
0=&&\int {\cal D}\Phi_H \frac {\delta}{\delta \Phi_H(p)}
e^{-S_{\L_0}(\Phi_S,\Phi_H)+J\Phi_H}\\=&&\int {\cal D}\Phi_H
\left[-D_H^{-1}\Phi_H -
\frac {\delta S^I_{\L_0}}{\delta \Phi_S}
+(-)^{\delta}J\right]e^{-S_{\L_0}(\Phi_S,\Phi_H)+J\Phi_H}\nonumber\\=&&
\left[-D_H^{-1}\frac {\delta W_{\L_R\L_0}}{\delta J}+(-)^{\delta}J+
\frac {\delta
W_{\L_R\L_0}}{\delta \Phi_S}+D_S^{-1}\Phi_S\right]e^{W_{\L_R\L_0}[J,\Phi_S]}
\enyn
where $\delta =0$ (or $1$) for (anti)commuting fields.
Making the Legendre transformation one obtains
\[
-D_H^{-1}\Phi_H^c+\frac{\delta \Gamma_{\L_R\L_0}}{\delta \Phi_H^c}+
D_S^{-1}\Phi_S -
\frac{\delta \Gamma_{\L_R\L_0}}{\delta \Phi_S} =0
\]
and using eq.\,(\ref{Gint}) one has
$$
\frac{\delta \bar\Gamma^{int}_{\L_R\L_0}}{\delta \Phi_H^c}-
\frac{\delta \bar\Gamma^{int}_{\L_R\L_0}}{\delta \Phi_S}=0
$$
which proves eq.\,(\ref{one}).
  
\subsection{Effective Ward identities}

The hard-soft decomposition is not gauge-invariant. As discussed in
\cite{Val} in the context of BRS quantization, the symmetry of the
original theory is not lost, but it is replaced by a non-local
symmetry on the hard and soft fields. Let us consider first the
simpler case of an abelian gauge theory, in which it is not necessary
to introduce ghosts.

Consider an abelian gauge theory with gauge-fixed
classical action $S_{cl}(\Phi)$ and infinitesimal gauge transformations 
\bey
\delta \Phi = R(\omega)\Phi + T(\omega) 
\eny
where $\omega$ is the infinitesimal gauge parameter.
For instance in electrodynamics the field content is
$ \Phi = ( A_\mu , \psi,\bar \psi )$ ,
with gauge transformations          
\[
\delta A_\mu = -{ 1\over e} \partial_\mu \omega\qquad
\delta \psi = i \omega \psi\qquad\delta \bar \psi = - i \omega \bar \psi
\]
The classical action transforms as
$\delta S_{cl}(\Phi)=c(\omega)\Phi$ 
 where $c \Phi$ is the breaking term due to the (linear covariant)
 gauge-fixing.

After the scale decomposition described above
the gauge symmetry acts non-locally on the hard and soft fields; under
\bea\label{HSG}
\delta \Phi_S = K_S\left[R (\Phi_S + \Phi_H) +T\right]\qquad
\delta \Phi_H = K_H\left[R (\Phi_S + \Phi_H) +T\right]
\ena
the action (\ref{action}) transforms as 
\bea\label{def1O}
\delta S_{\L_0}(\Phi_S,\Phi_H)=c ( \Phi_S +\Phi_H) + {\cal O}_{\L_0}
(\Phi_S ,\Phi_H;\omega)
\ena
${\cal O}_{\L_0}$ 
depends
on its arguments only through their sum; at tree
level
it is an irrelevant term vanishing for
$\L_0 \rightarrow \infty$.

Multiplying eq.\,(\ref{def1O}) by $e^{-S_{\L_0}(\Phi_S,\Phi_H)+J\Phi_H}$ and 
performing a functional
integration over $\Phi_H$ , after an integration by parts one gets:
\bea\label{id1O}
0 = \int {\cal D}\Phi_H\left[\delta\Phi_S\frac{\delta}{\delta\Phi_S} 
- J\delta\Phi_H +c ( \Phi_S +\Phi_H)+ {\cal O}_{\L_0}(\Phi_S +\Phi_H;\omega)\right]
e^{-S_{\L_0}(\Phi_S,\Phi_H)+J\Phi_H}
\ena
The correctness of  the naive procedure leading to eq.\,(\ref{id1O})
is ensured, in the perturbative framework, by the Quantum Action
Principle \cite{Qap,Becchi1};
if there are no
anomalies ${\cal O}_{\L_0}$ is evanescent at quantum level, being
evanescent at tree level. 
We will prove this point in the third section using the flow equation
technique.

Making the Legendre transformation to $\Gamma_{\L_R \L_0}$, using 
eqs.\,(\ref{Gint},\ref{one})
and collecting terms, the effective Ward identity depends only on the field
$\Phi = \Phi_S + \Phi_H^c$; one gets
\bea\label{ward}
(R\Phi +T)
\left[K_{\L_0}\frac{\delta\Gamma^{int}_{\L_R\L_0}}{\delta\Phi}+
D^{-1}\Phi \right]- c\Phi =
{\cal T}_{\L_R\L_0}[\Phi;\omega]+{\cal O}_{\L_R\L_0}[\Phi;\omega]
\ena
where ${\cal O}_{\L_R\L_0}[\Phi;\omega]$ is the functional generator
of the operator insertion corresponding 
to ${\cal O}_{\L_0}$ and
\bey\label{tau}
{\cal T}_{\L_R\L_0}[\Phi;\omega] \equiv
tr K_S R {\delta^2 W_{\L_R \L_0} \over \delta J \delta \Phi_S } =
tr D_H^{-1}K_{\L_R}R 
{\left[D_H^{T-1}+
\frac{\delta^2\Gamma^{int}_{\L_R\L_0}}{\delta\Phi^2}\right]}^{-1}+ const.
\eny
in which the trace includes a
momentum loop and 
${\left[\frac{\delta^n\Gamma}{\delta\Phi^n} \right]}_{i_1...i_n}\equiv
\frac{\delta^n\Gamma}{\delta\Phi_{i_1}... \delta\Phi_{i_n}}$;\ 
$const.$ is an unimportant field-independent term.

${\cal T}_{\L_R\L_0}$ is the non-linear part of the
effective Ward identity on $\Gamma_{\L_R\L_0}$; it has the form of a one-loop
skeleton diagram.
If the proper vertices are renormalized up to loop $l-1$,
 then ${\cal T}_{\L_R\L_0}$ is
finite at loop $l$; in fact the loop contained in the trace 
has the ultraviolet
cut-off $K_{\L_R}$, so that 
${\cal T}_{\L_R\L_0}\rightarrow {\cal T}_{\L_R}$
finite for $\L_0\rightarrow\infty$.
If there are no anomalies, it is possible to choose 
 $\Gamma_{\L_R\L_0}|_{ rel}$
such that ${\cal O}_{\L_R\L_0}[\Phi;\omega]$ is evanescent, namely beyond the
tree level:
\bey\label{ren}
\Big[(R\Phi +T) \frac{\delta\Gamma^{int}_{\L_R\infty}}
{\delta\Phi}\Big]{|_{ rel}}=
(R\Phi +T) \frac{\delta\Gamma^{int}_{\L_R\infty}|_{ rel}}{\delta\Phi}=
{\cal T}_{\L_R\infty}|_{ rel}[\Phi;\omega]
\eny
Since $\L_R > 0$, the 1PI Green functions are regular functions 
of the momenta
(in particular in the origin). Therefore  in (\ref{ren}) the relevant
terms 
can be defined at zero-momentum in the hard-soft renormalizaton scheme
and the first equality is actually  trivial.

In sect.\,III we will show that if eq.\,(\ref{ren}) is satisfied the
effective Ward identities of the
hard-soft theory become exact in the limit $\L_0 \to \infty$:
\bea\label{wardex}
(R\Phi +T)
\left[\frac{\delta\Gamma^{int}_{\L_R\infty}}{\delta\Phi}+
D^{-1}\Phi \right]- c\Phi =
{\cal T}_{\L_R}[\Phi;\omega]
\ena
In Sect. III it will be shown that, if eq.\, (\ref{ren}) holds, then
the usual Ward identity on $\Gamma_{0\infty}[\Phi]$ is satisfied.
  
The effective Ward identities for a composite operator ${\cal O}$ can be
obtained
easily by introducing a source term $\int\, dx \eta(x){\cal O}(x)$ in
the action (\ref{action}), and then by differentiating with respect
to $\eta(x)$ in $\eta=0$ the extended version of eq.\,(\ref{ward}).
We will consider the case of a gauge-invariant definition of
the composite operator $i\bar{\psi}(x)\gamma_\mu\gamma_5\psi(x)$
with an associated source $\eta_{5\mu}(x)$. The action with this
source
term is invariant at tree level under the local axial transformation:
\bey
&&\delta_5 A_\mu(x) =0 \qquad\qquad\qquad
\delta_5\psi(x)=i\omega_5(x)\gamma_5\psi(x)\nonumber\\
&&\delta_5\bar{\psi}(x)=i\omega_5(x) \bar{\psi}(x)\gamma_5\qquad
\delta_5\eta_{5\mu}(x)= - \partial_\mu\omega_5(x) 
\eny
In a compact notation we shall write the previous formula 
$\delta_5\Phi=R_5 \Phi$ and $\delta_5\eta_5=T_5$
Proceeding as in eqs.\,(\ref{id1O},\ref{ward}) one arrives at
\bey\label{ward5}
&&(R_5\Phi +T_5)
\left[K_{\L_0}\frac{\delta\Gamma^{int}_{\L_R\L_0}}{\delta\Phi}+
D^{-1}\Phi \right]
+T_5\frac{\delta\Gamma^{int}_{\L_R\L_0}}{\delta\eta_5}=\\&&=
tr D_H^{-1}K_{\L_R}R_5 
{\left[D_H^{T-1}+
\frac{\delta^2\Gamma^{int}_{\L_R\L_0}}{\delta\Phi^2}\right]}^{-1}
+{\cal O}_{5\L_R\L_0}[\Phi;\omega]\nonumber
\eny
where ${\cal O}_{5\L_R\L_0}[\Phi;\omega]$ is evanescent at tree level, but
not
at quantum level as we will see in sections II and III.

The fact that in the hard-soft decomposition the symmetries of the
original lagrangian are not lost, but become non-local symmetries
which are well-defined at the quantum level has been first shown
in \cite{Val}; it is true not only in the abelian but also in the non-abelian case.

Consider an action $S_{\L_0}(\Phi)$ of the form (\ref{act}) 
which modulo irrelevant 
terms is invariant under the BRS transformations $\delta \Phi =
\epsilon P(\Phi)$;
$\Phi$ denotes physical and ghost fields, $\epsilon$ is an
anticommuting parameter and $P(\Phi)$ is polynomial in the fields.
After the hard-soft decomposition the BRS symmetry
is replaced by 
the non-local BRS transformations \cite{Val}:
\bea\label{BRS}
\delta \Phi_S =\epsilon K_S P(\Phi_S + \Phi_H) \qquad\qquad
\delta \Phi_H =\epsilon K_H P(\Phi_S + \Phi_H)
\ena
which is analogous to (\ref{HSG}).
The action (\ref{action}) transforms as 
\bey\
\delta S_{\L_0}(\Phi_S,\Phi_H)= \epsilon {\cal O}_{\L_0}
(\Phi_S +\Phi_H)
\eny
where ${\cal O}_{\L_0}$ is evanescent at tree
level.
Adding to $J\Phi_H$ the source term $\eta P(\Phi_S +\Phi_H)$ and 
proceeding as in eqs.\,(\ref{id1O},\ref{ward},\ref{tau}) we obtain:
\bea\label{Slav}
-\frac{\delta\Gamma^{int}_{\L_R\L_0}}{\delta\eta}
\left[K_{\L_0}\frac{\delta\Gamma^{int}_{\L_R\L_0}}{\delta\Phi}+
D^{-1}\Phi \right] =
{\cal T}_{\L_R\L_0}[\Phi,\eta]+{\cal O}_{\L_R\L_0}[\Phi,\eta]
\ena
where ${\cal O}_{\L_R\L_0}[\Phi,\eta]$ is the functional generator
of an operator insertion which is irrelevant at tree level and
\bey
{\cal T}_{\L_R\L_0}[\Phi,\eta] \equiv
tr K_S  {\delta^2 W_{\L_R \L_0} \over \delta \eta \delta \Phi_S } =
 -tr D_H^{T-1}K_{\L_R} \frac{\delta^2\Gamma^{int}_{\L_R\L_0}}
{\delta\eta\delta\Phi}
{\left[D_H^{T-1}+
\frac{\delta^2\Gamma^{int}_{\L_R\L_0}}{\delta\Phi^2}\right]}^{-1}
\eny
Eq.\,(\ref{Slav}) is  the
effective Slavnov-Taylor identity which has been derived in the
context of the exact renormalization flow in \cite{Becchi2,El,DM}.

Let us finally remark on the choice of the cut-off function.
In \cite{LM,Val} the cut-off $K_{\L} = {\L^2 \over p^2 + \L^2}$ 
is used.
This cut-off does not eliminate completely the divergences in the soft
theory, which has soft propagator and Wilsonian vertices; loops containing
a single soft propagator can be divergent;
renormalization conditions on the two-point functions must be
imposed after integrating over $\Phi_S$.
The non-local BRS transformations introduced in \cite{Val} have not
the same form as eq.\,(\ref{BRS}); however in the Feynman gauge they
coincide with the $\L_0 \to \infty$ limit of eq.\,(\ref{BRS}) provided
the above-mentioned cut-off function is chosen.

\section{QED at one loop}
\subsection{Ward identities at one loop}
Let us illustrate a hard-soft renormalization scheme in the case of massless
QED at one loop. 
At tree level the action is, in the Feynman gauge,
\bey
S^{(0)}=\int_p  {1\over2}A_\mu(-p) p^2 A_\mu(p)+
\bar{\psi}(-p)i\ds{p}\psi(p) + 
\int_{p_1p_2}\bar{\psi}(p_1) ie\gamma^\mu\psi(p_2)A_\mu(-p_1-p_2)
\eny
where $\int_p \equiv \int {d^4p \over (2\pi)^4}$.
The propagators are
\bey
S(p) = {-i \over \ds{p} } \qquad  \qquad  D_{\mu\nu}(p) = 
{1 \over p^2}\delta_{\mu\nu}
\eny
To renormalize the theory in the HS scheme it is sufficient to
renormalize the Wilsonian theory with hard propagators $D_H$
(see eq.\,(\ref{prop})) and bare interacting action, whose $l$-th term
in the loop expansion is 
\bey\label{sl}
S^{(l)I}_{\L_0}=\int_p {1\over2}A_\mu(-p)\left[c^{(l)}_1\delta_{\mu\nu}+
c^{(l)}_2(p^2\delta_{\mu\nu}-p_\mu
p_\nu)+c^{(l)}_3p^2\delta_{\mu\nu}\right]A_\nu(p)+
c^{(l)}_4 \bar{\psi}(-p)i\ds{p}\psi(p)+\nonumber\\
\int_{p_1p_2} c^{(l)}_5 \bar{\psi}(p_1)
ie\gamma^\mu\psi(p_2)A_\mu(-p_1-p_2)+\int_{p_1p_2p_3}{c^{(l)}_6 \over 8} 
A_\mu(p_1)A_\mu(p_2)A_\nu(p_3)A_\nu(-\sum_{i=1}^3p_i)
\eny
 The renormalization 
conditions are chosen loop by loop imposing eq.\,(\ref{ren});
they depend on two  renormalization constants and on the cut-off.
To be able to determine these conditions analytically we will 
choose a simple cut-off function: 
\bea\label{cutoff}
K_{\L}(p) = {\L^4 \over (p^2 + \L^2 )^2}
\ena

The usual Green functions are then computed using the previously determined 
bare action and the propagator $D K_{\L_0}$.
As a consequence of eq.\,(\ref{Z}), this is equivalent to computing 
the Green functions with the
action (\ref{action}), $S^I_{\L_0}$ being given in (\ref{sl}), with 
 $ A_\mu = A_{S\mu}+A_{H\mu}$ and $\psi= \psi_S + \psi_H $.

 At one loop the effective Ward identity (\ref{ward}) on the photon two-point function is
\bey\label{Wone}
{1 \over e}p_\nu\Gamma_{\nu\mu}^{(1)\L_R\infty}(p)=
{\cal T}_{\mu}^{(1)\L_R\infty}(p)
\eny
where
\bey\label{Wonea}
{\cal T}_{\mu}^{(1)\L_R\infty}(p)=
2ie\int_qK_{\L_R}(q) TrS^{\L_R\infty}(q+p)\gamma_\mu
\eny
Using the Lorentz invariant decomposition
$
\Gamma_{\mu\nu}(p)=A(p^2)(p^2\delta_{\mu\nu}-p_\mu p_\nu)+
B(p^2)\delta_{\mu\nu}
$  
the renormalization conditions on the photon two-point function
compatible with the effective Ward identities are
\bey\label{renph}
A^{(1)\L_R \infty} (0) = z^{(1)}_3 \qquad B^{(1)\L_R \infty} (0) =
 {5 \over {24}} {{e^2 \L_R^2} \over {\pi^2}} \qquad
\partial_{p^2} B^{(1)\L_R \infty} (p^2)|_{p^2=0} =
- {e^2 \over 24 \pi^2}  
\eny
The renormalization conditions on the electron two-point function
and on the electron-photon vertex are
\bea\label{RCE}
\Sigma^{(1)\L_R\infty}(0)=0\;  \qquad 
\frac{\partial \Sigma^{(1)\L_R\infty}}{\partial p_\mu}(p)|_{p=0}=
i\gamma_\mu z^{(1)}_2 \; \qquad
\Gamma^{(1)\L_R\infty}_\mu(0,0)=ie\gamma_\mu z^{(1)}_1 \; 
\ena
(the first condition is required by the rigid axial invariance).\\
Using the effective Ward identity (\ref{ward}) we obtain (see also \cite{bon1})
\bea\label{Wtwo}
{1 \over
e}(p-q)_\mu\Gamma_{\mu}^{(1)\L_R\infty}(p,-q)+\Sigma^{(1)\L_R\infty}(q)-
\Sigma^{(1)\L_R\infty}(p)=
{\cal T}^{(1)\L_R\infty}(p,-q)
\ena
where
\bea\label{Wtwoa}
{\cal T}^{(1)\L_R\infty}(p,-q)=
-e^2\int_l K_{\L_R}(l)\left[\gamma_\mu S^{\L_R\infty}(p-q+l)
\gamma_\nu D_{\mu\nu}^{\L_R\infty}(l-q)-
(q\leftrightarrow p)\right]
\ena
The renormalization conditions (\ref{RCE}) are compatible with the
effective Ward  identity (\ref{Wtwo}) provided
\bey
i\gamma_\mu (z^{(1)}_2 - z^{(1)}_1) =
-{1 \over e}\frac{\partial {\cal
T}^{(1)\L_R\infty}(p,0)}{\partial p_\mu}|_{p=0}=i
\gamma_\mu {53\over 960}{e^2\over\pi^2}
\eny
The last non-vanishing renormalization condition is determined by the
effective Ward identities to be (see also \cite{bon1})
\bea\label{RCL}
\Gamma_{\mu\nu\rho\sigma}^{(1)\L_R \infty}(0,0,0) = 
- {e^4 \over 12 \pi^2}
\left(\delta_{\mu\nu} \delta_{\rho\sigma} + \delta_{\mu\rho} \delta_{\nu\sigma}
+ \delta_{\mu\sigma} \delta_{\nu\rho} \right)
\ena
Having  fixed the renormalization conditions, we can determine the
counterterms in the bare action
\bey
c_1^{(1)}= {e^2 \L_0^2\over 24 \pi^2 }\qquad\qquad\qquad\qquad\qquad\ 
&&c_2^{(1)}= {e^2 \over 12 \pi^2} (ln{\L_R^2 \over \L_0^2} + {19 \over 15}) + 
z_3^{(1)}\quad\ \                  
c_3^{(1)}= - {e^2 \over 24 \pi^2}\nonumber\\
c_4^{(1)}= {e^2 \over 16 \pi^2} (ln{\L_R^2 \over \L_0^2} + {51 \over 20})
+ z_1^{(1)}\qquad  &&c_5^{(1)}= {e^2 \over 16 \pi^2} (ln{\L_R^2 \over \L_0^2} + {5 \over 2})
+ z_1^{(1)} \qquad\ 
c_6^{(1)}= - {e^4 \over 12 \pi^2}
\eny
These counterterms include the finite parts needed to satisfy the
effective Ward
identities at $\L_R$; they are easily computed due to the choice of
the HS scheme and of the cut-off (\ref{cutoff}).
A partial determination of the one-loop counterterms in QED using
the exact renormalization group approach  and renormalization 
conditions at $\L=0$
can be found in \cite{bon1} and
\cite{BS}; in \cite{BS} the
cut-off  (\ref{cutoff}) is also used.

After integrating out the soft modes, the usual Ward
identities are automatically satisfied. Let us check them in two
cases.

The vacuum polarization is transverse:  
\bey
A^{(1)0\infty}(p^2)  = {e^2 \over 12 \pi^2} 
(- ln {p^2 \over \L_R^2} + {7 \over 15} ) + z_3^{(1)} \qquad \qquad  \qquad
B^{(1)0 \infty}(p^2) = 0
\eny

The self-energy of the electron
\bey
\Sigma^{(1)0 \infty}(p)  = i \ds{p} \left[ {e^2 \over 16 \pi^2} 
(- ln {p^2 \over \L_R^2} + {103 \over 60} ) + z_1^{(1)} \right]
\eny
and the vertex
\bey
\Gamma^{(1)0\infty}_{\mu}(p,-q)  =&& i e \gamma_{\mu} \left[ z_1^{(1)}- 
{e^2 \over 8\pi^2} \left( {77 \over 120}+\int_0^1 dx \int_0^{1-x} dy\; 
ln {Q^2 \over \L_R^2}  \right) \right]\nonumber\\&&+
{i e^3 \over 8\pi^2} \int_0^1 dx \int_0^{1-x} dy
{[-\ds{ p} x +\ds{ q} (1-y) ] \gamma_{\mu} [- \ds{q} y
+ \ds{p} (1-x) ] \over Q^2}\\
Q^2 =&& p^2 x(1-x) + q^2 y(1-y) - 2p \cdot qxy \nonumber
\eny 
satisfy the relation:
$ (p-q)_\mu\Gamma^{(1)0\infty}_{\mu}(p,-q)+e[\Sigma^{(1)0 \infty}(q) -
\Sigma^{(1)0 \infty}(p)]=0 $.

\subsection{Axial Ward identities at one loop}

The effective axial Ward identities can be discussed
analogously to the effective gauge Ward identities. The HS renormalization
conditions are chosen compatible with the effective gauge Ward
identities; we will verify now that it follows that the effective
axial Ward identities are anomalous; only zero-momentum graphs at
$\L_R$ need to be evaluated to determine the anomaly.
 For a computation of the axial anomaly in QED using the flow equation
and imposing the usual gauge Ward identity  see \cite{bon1}.

 At one loop the fermion-axial vector vertex is equal to the fermion-photon
vertex, multiplied by $ \gamma_5$; the same is true for the corresponding
effective axial Ward identity, which is equivalent to the effective 
gauge Ward identity eq.\,(\ref{Wtwo}).

The matrix element of  $i\bar{\psi} \gamma^{\mu} \gamma_5 \psi (x)$ in
the two-photon sector must include a counterterm due to the possible
mixing
with  $\epsilon_{\mu \nu\rho \sigma}F^{\nu \rho} A^{\sigma}$:
\bey\label{AVV}
\Gamma_{ 5 \mu, \nu \rho}^{(1)\L_R \L_0}(q,p) = &&
2i e^2 \int_l Tr \gamma_{\mu} \gamma_5 S^{\L_R \L_0}(l-q) \gamma_{\nu}  
S^{\L_R \L_0} (l) \gamma_{\rho} S^{\L_R \L_0}(l + p)\nonumber\\&& +\hbox{}
c^{(1)} \epsilon_{\mu \nu \rho \sigma} (p - q)_{\sigma}
\eny
It satisfies the gauge effective Ward identity (\ref{ward})
\FL
\bey\label{WAone}
q_{\nu} \Gamma_{5 \mu, \nu \rho}^{(1)\L_R\L_0}(q,p) =&& 
2 e^2 \int_l Tr \gamma_{\mu} \gamma_5 \left[ K_{\L_R}(l-q) S^{\L_R \L_0}(l)-
S^{\L_R \L_0}(l-q) K_{\L_R}(l) \right] \gamma_{\rho} S^{\L_R
\L_0}(l+p)\nonumber\\ 
&&+\hbox{} {\cal O}_{5 \mu \rho}^{(1)\L_R \L_0}(q,p)
\eny
and the axial effective Ward identity (\ref{ward5})
\FL
\bey\label{WAtwo}
(q+p)_{\mu} \Gamma_{5 \mu, \nu \rho}^{(1)\L_R \L_0}(q,p) =&& \nonumber
2 e^2 \int_l Tr \gamma_{\rho}\gamma_5 \left[ K_{\L_R}(l-q) S^{\L_R\L_0}(l+p)-
(p \leftrightarrow -q) 
\right] \gamma_{\nu} S^{\L_R \L_0}(l)\\ &&+\hbox{} 
\widetilde{\cal O}_{ 5 \nu \rho}^{(1)\L_R \L_0}(q,p)
\eny
The renormalization conditions on $\Gamma_{ 5 \mu, \nu \rho}^{(1)\L_R \L_0}(q,p)$
are chosen such that ${\cal O}_{5 \nu \rho}^{(1)\L_R \L_0}$ is evanescent:
$
\frac{\partial^2}{\partial q_{\alpha} \partial p_{\beta} }
{\cal O}_{ 5 \nu \rho}^{(1)\L_R \infty}(q,p)|_{p=q=0} = 0
$
which implies
\bey\label{div}
\frac{\partial}{\partial p_{\sigma} }
\Gamma_{ 5 \mu, \nu \rho}^{(1)\L_R \infty}(0,p) |_{p=0} =&& 
4 e^2 \int_l Tr \frac{\partial K_{\L_R}}{\partial l^{\nu}} \gamma_5 
\gamma_{\mu} S^{\L_R \infty} \gamma_{\rho}
\frac{\partial S^{\L_R \infty}}{\partial l^{\sigma}}  \nonumber\\
=&&\frac{4 e^2}{3} \epsilon_{\mu \nu \rho \sigma} 
\int_l \frac{\partial}{\partial l_\alpha}\left[
\left(1-K_{ \L_R}\right)^{3}  \frac{l^{\alpha}}{(l^2)^2} \right]=
\frac{e^2}{6 \pi^2}
\epsilon_{\mu \nu \rho \sigma}
\eny
This condition determines the value of $c^{(1)}$ of eq.\,(\ref{AVV}).
It follows that $ \widetilde {\cal O}$ is fixed by the effective axial
Ward identity eq.\,(\ref{WAtwo}) to satisfy
\bea
\frac{\partial^2}{\partial q_{\alpha} \partial p_{\beta} }
\widetilde{\cal O}_{ 5 \nu \rho}^{(1)\L_R \infty}(q,p)|_{p=q=0} =
\frac{e^2}{2 \pi^2}
\epsilon_{\nu \rho \alpha \beta}
\ena
which is the anomaly of the axial Ward identity.
Observe that in this formulation the anomaly is evaluated in terms of 
explicitly
zero momentum finite integrals, which do not depend on the explicit
form of the cut-off function. In fact the integral in (\ref{div}) is
a total divergence.
It is interesting to observe that the triangle graph does not give any
contribution to the anomaly for $\L_R \neq 0$, so that
$c^{(1)}=\frac{e^2}{6 \pi^2}$; in fact
\bea\label{sur}
\frac{\partial}{\partial p_{\beta} }
\Gamma_{ 5 \mu, \nu \rho}^{(1)\L_R \L_0}(0,p) |_{p=0} =
- \frac{8 e^2}{3} \epsilon_{\mu \gamma \rho \delta} 
\int_l \frac{\partial}{\partial l_\beta}\left[\left( K_{\L_0}-K_{ \L_R}\right)^{3} (l) \frac{\partial}{\partial l_{\nu} }
\left( \frac{l^{\gamma}}{l^2} \right) \frac{l^{\delta}}{l^2} \right] = 0
\ena
 For example in the limit in which 
$K_\L$ is the step function 
the integral in eq.\,(\ref{sur}) is proportional to
\bea
\left( \int_{\Sigma_{\L_0}} d \Sigma_{\beta} - 
\int_{\Sigma_{\L_R}} d \Sigma_{\beta} \right) 
\frac{\partial}{\partial l_{\nu} }
\left( \frac{l^{\gamma}}{l^2} \right) \frac{l^{\delta}}{l^2} = 0
\ena
The two surface contributions, coming from the spheres $l^2 = \L_0^2$
and $l^2 = \L_R^2$ cancel each other.

The $AAA$ vertex satisfies at one loop an effective Ward identity which 
is formally the same as eq.\,(\ref{WAtwo}), but now the vertex 
$\Gamma_{\mu \nu \rho}^{\L_R \L_0}$ is completely symmetric, so that it 
admits no counterterm.
Therefore $\widetilde{\cal O}^{rel} = -{\cal T}^{rel}$, and the $AAA$ anomaly
 is one-third of the $AVV$ anomaly.

\section{Exact renormalization group}
In the first section we considered a fixed scale decomposition, in which
the soft and hard modes are separated at the renormalization scale
$\Lambda_R$. Changing the separation point to an arbitrary 
scale $\Lambda >0$ and changing
appropriately the renormalization conditions, i.e. choosing the
parameters $z_{1 \Lambda}$ and $z_{3 \Lambda}$, which are not constrained
by the effective Ward identities, the physical quantities are unchanged.
It is possible to write an exact renormalization flow equation describing
the continuous change of the Wilsonian effective action 
$W_{\Lambda \Lambda_0}$.

The generating functional of the connected (interacting)
amputated Green functions
$L_{\Lambda\L_0}[\phi,\Phi_S]$ satisfies 
$$
W_{\Lambda\L_0}[J,\Phi_S] ={ 1 \over 2}JD^T_{\L\L_0}J-
{ 1 \over 2}\Phi_S D_{0\L}^{-1}\Phi_S
-L_{\Lambda\L_0}[D^T_{\L\L_0}J,\Phi_S]
$$
The same proof which led to eqs.\,(\ref{Gint},\ref{one}) shows 
that actually $L_{\Lambda\L_0}$
depends on its two arguments only through their sum $\Phi_S +\phi \equiv
\Phi$ so we will consider the simpler functional $L_{\Lambda\L_0}[\Phi]$
(see also \cite{Morris}).
Another way to see this fact exploits the following representation of
the
functional $L_{\Lambda\L_0}$ \cite{KKS}:
\bea\label{fluxintL}
e^{-L_{\L\L_0}[\phi +\Phi_S]} = e^{\Delta_{\Lambda\Lambda_0}
\left({\delta \over\delta \phi}\right)}
e^{-L_{\L_0}[\phi+\Phi_S]}
\ena
where $L_{\Lambda_0}$ is equal to the
bare lagrangian, apart from the tree-level kinetic term, and 
$ \Delta _{\Lambda\Lambda_0}$ is the functional Laplacian:
$$
\Delta_{\Lambda\Lambda_0}
= { 1 \over 2}{\delta \over\delta \phi_i}
D^{\Lambda\Lambda_0}_{ji}{\delta \over\delta \phi_j}
$$
$L_{\Lambda_0}$ depends on $\phi$ and on the background $\Phi_S$
through
their sum, therefore  $L_{\Lambda\L_0}$ will depend only on $\Phi_S
+\phi$.
The above proof shows that, in case of propagators $D_{\L\L_0}$ with
compact
support, the amputated Green functions of the theory, a priori defined
only for momenta in this support, can be continued to arbitrary
momenta by performing functional derivatives with respect to  
the background field.

$ L_{\L\L_0}[\Phi]$ satisfies the flow equation \cite{Polchi}
\bea\label{flux}
\partial_\Lambda L_{\Lambda\L_0} =
\left(\partial_\Lambda \Delta_{\Lambda
\Lambda_0}\right)L_{\Lambda\L_0} -{ 1 \over 2}
L'_{\Lambda\L_0}\left(\partial_{\L}D^T_{\Lambda\Lambda_0}\right)
L'_{\Lambda\L_0}
\ena
where $L'_{\Lambda\L_0} \equiv {\delta L_{\Lambda\L_0}\over\delta \Phi}$.

In the Polchinski approach \cite{Polchi} $L_{\Lambda_0}$ is the boundary
condition
for the flow equation (\ref{flux}). The coefficients of the relevant
polynomials 
in the fields, which appear in $L_{\L_0}$, must depend in a 
suitable way on the
ultraviolet cut-off.
This dependence is fixed by imposing the renormalization conditions
 on the relevant components of the functional 
$L_{\Lambda_R\L_0}$.
Perturbatively the relations between the renormalization conditions
and the bare boundary condition is invertible and no
ambiguities are involved.

Using a cut-off function with compact support, Polchinski gave a
simple proof of
the power-counting renormalization theorem; the proof has been
generalized and further simplified in \cite{KKS,KKop,KKop2,KK}.
Analogous results have been obtained using an exponential
cut-off in \cite{BT};
a proof of renormalizability with a Pauli-Villars cut-off, like that
of eq.\,(\ref{cutoff}), has not yet been given using only the flow
equation.

An expression analogous to eq.\,(\ref{fluxintL}) holds for the
 generating
 functional of
the connected and amputated graphs with 
an insertion of a composite operator ${\cal M}$:
\bea\label{fluxintO}
{\cal M}_{\L\L_0}[\phi,\Phi_S] =  e^{L_{\L\L_0}[\phi,\Phi_S]}
e^{\Delta_{\Lambda\Lambda_0}
\left({\delta \over\delta \phi}\right)}{\cal M}_{\L_0}[\phi,\Phi_S]
e^{-L_{\L_0}[\phi,\Phi_S]}
\ena
where ${\cal M}_{\L_0}$ is the bare insertion.
If ${\cal M}_{\L_0}$, as well as $L_{\L_0}$, depends on $\phi +\Phi_S$
eq.\,(\ref{fluxintO}) shows that also ${\cal M}_{\L\L_0}$ depends on
 the
fields only through 
$ \Phi$.
${\cal M}_{\Lambda \Lambda_0}[\Phi]$ satisfies, by construction, the 
linear differential equation for the connected insertion of 
an operator:
\bea\label{fluxO}
\partial_\Lambda{\cal M}_{\L\L_0}=
\left(\partial_\L \Delta_{\L\L_0}\right){\cal M}_{\L\L_0}
- L'_{\L\L_0}\left(\partial_{\L} D^T_{\L
\L_0}\right){\cal M}^{\,'}_{\L\L_0}
\ena

Eq.\,(\ref{fluxO}) is the starting point to prove the renormalizability of a
composite operator of dimension $d$, defined by a bare boundary
condition ${\cal M}_{\L_0}$ at $\L=\L_0$ (a polynomial of dimension
$d$ 
in the fields,
compatible with the rigid symmetries) and by the renormalization
conditions
at $\L=\L_R$. In particular if at $\Lambda = \Lambda_R$ 
the relevant part of ${\cal M}_{\Lambda_R \Lambda_0}$ is zero (or suitably
vanishing as $\Lambda_0 \rightarrow \infty$) and if the irrelevant part
of ${\cal M}_{\Lambda_0 }$ fulfils suitable 
bounds in the $\Lambda_0$ 
dependence then
$lim_{\Lambda_0 \to \infty} {\cal M}_{\Lambda \Lambda_0} = 0$ 
\cite{KKop}.

\subsection{Effective Ward identities and infrared convergence}

The validity of the Ward identities in QED has been studied using
the flow equation in \cite{KKQED1,KKQED2} choosing 
renormalization conditions at $\L = 0$ compatible with the usual Ward 
identities.

We want to show that, choosing a HS scheme satisfying
eq.\,(\ref{ren}), the exact effective Ward identities (\ref{wardex}) are
satisfied; moreover the usual Ward identities and the infrared finiteness 
of the theory follow without further constraints for $\L \to 0$.

The effective Ward identity on $L_{\Lambda\L_0}$ is obtained defining
\bey\label{defO}
{\cal O}_{\L\L_0}[\Phi;\omega]\ &&=
i\int dx \omega(x){\cal O}_{\L\L_0}[x,\Phi] =\\&&= e^{{L_{\L\L_0}}[\Phi]} 
e^{\Delta_{\L\L_0}\left(\frac
{\delta}{\delta\Phi}\right)}\left\{-\left[(R\Phi +T)K_{\L_0}L'_{\L_0}[\Phi] +
\Phi D^{-1}R\Phi\right]e^{-L_{\L_0}[\Phi]}\right\}\nonumber
\eny
and proving that ${\cal O}_{\L\L_0} \to 0 $ for $\L_0 \to \infty$.

$\Phi$ is the multiplet of independent fields 
$( A_{\mu}, \psi, \bar{\psi} )$ and $R$ and $T$ are the parameters
describing the gauge transformations, introduced in the first section.
Notice that the argument in square brackets on the r.h.s. of this
equation is equal to ${\cal O}_{\L_0}[\Phi;\omega]$ 
defined in eq.\,(\ref{def1O}), with $\Phi = \Phi_S + \Phi_H$.

One wants to show that it is possible to choose the renormalization
conditions on $ L_{\L_R\L_0}$ in such a way that  the
operator
defined in eq.\,(\ref{defO}) is evanescent.
Indeed the relevant parts of ${\cal O}_{\Lambda_R \Lambda_0}$ can be
related to those of $L_{\Lambda_R\L_0}$ by the interpolating effective
Ward identity
which can be obtained simply by commuting the argument in the square brackets
in (\ref{defO}) with the exponential of the functional Laplacian:
\bey\label{idO}{\cal O}_{\L\L_0}[\Phi;\omega] = 
&&-TK_{\L_0}L^{'}_{\L\L_0}[\Phi]
-\Phi D^{-1}R\Phi
+\Phi D^{-1}RD_{\L\L_0}L^{'}_{\L\L_0}[\Phi]\\&&
-\Phi R^TK_{\L}L^{'}_{\L\L_0}[\Phi]
+L_{\L\L_0}^{'}[\Phi]D^T_{\L\L_0}R^TK_{\L} L^{'}_{\L\L_0}[\Phi]
-{\cal T}_{\L\L_0}[\Phi;\omega]\nonumber
\eny
where
\bea\label{defT}
{\cal T}_{\L\L_0}[\Phi;\omega]=
i\int dx \omega(x){\cal T}_{\L\L_0}[x,\Phi]= 
trK_\L RD_{\L\L_0} L^{''}_{\L\L_0} [\Phi]
\ena

We realize that only the first three terms in the r.h.s. of 
eq.\,(\ref{idO}) are present in the broken Ward identity of 
ref.\cite{KKQED1},\cite{KKQED2} all the
others terms being
contained in a redefinition of ${\cal O}_{\Lambda \Lambda_0}$, which
then does not
satisfy the linear equation of the insertions (\ref{fluxO}).
 These three addenda are the
only ones surviving in the formal limit $\Lambda \rightarrow 0$,
$\Lambda_0 \rightarrow \infty$, giving the usual Ward identities.

From now on we consider explicitly the case of the cut-off function
$K\left({p^2 \over \L^2}\right)$ with \hbox{$K = K(x)$} a $C^{\infty}$ function
with compact support (equal to $1$ for $x \leq 1$ and equal to zero for
$x \geq 4$).

In order to connect the relevant parts of $L_{\Lambda_R\L_0}$ and ${\cal
O}_{\Lambda_R\L_0}$ it is sufficient to consider the functional
derivatives of eq.\,(\ref{idO}) with respect to the fields
$\bar{\psi}_{\alpha_1}(p_1)\ldots\psi_{\beta_n}(q_n)\ldots
A_{\mu_m}(k_m)$
and to the gauge parameter $\omega$, 
in a suitable neighbourhood of the origin of
momentum space. Noticing that the terms 
$\Phi D^{-1}RD_{\Lambda\L_0}L^{'}_{\L\L_0}[\Phi]$
and $L_{\L\L_0}^{'}[\Phi]D_{\Lambda\L_0}R^TK_{\L}
L^{'}_{\L_0\L}[\Phi]$ 
do not give
contribution
in this region, we obtain for the $l$-th term in the loop expansion:
\bey\label{compO} 
&&{\cal O}^{(l)\L\L_0 }_{2n\,m}
(\alpha_1p_1,\ldots , \alpha_np_n;\beta_1q_1,\ldots, \beta_nq_n;
\mu_1k_1,\ldots ,\mu_mk_m)= \nonumber\\
&&{1\over e}P_{\mu}L^{(l)\L\L_0}_{2n\,m+1}
(\alpha_1p_1,\ldots , \alpha_np_n;\beta_1q_1,\ldots,
\beta_{n-1}q_{n-1},\beta_n;
\mu_1k_1,\ldots ,\mu_mk_m,\mu P)+\\
&&\sum_{j=1}^{n}(-1)^{n-j}\left[
L^{(l)\L\L_0}_{2n\,m}
({\cal P}_j)
 -L^{(l)\L\L_0}_{2n\,m}({\cal Q}_j)\right]-{\cal T}^{(l)\L\L_0}_{2n\,m}(\alpha_1p_1,..,\alpha_np_n;\beta_1q_1,..,
\beta_nq_n;\mu_1k_1,..,\mu_mk_m)\nonumber
\eny
where the multi-indices ${\cal P}_j$ and ${\cal Q}_j$ are defined by:\\
$ {\cal P}_j \equiv (\alpha_1p_1,..,\alpha_{j-1}p_{j-1},\alpha_{j+1}p_{j+1},..,
 \alpha_np_n,\alpha_j\,P\!+\!p_j;
 \beta_1q_1,..,\beta_{n-1}q_{n-1},\beta_n;\mu_1k_1,..,\mu_mk_m) $\\
$ {\cal Q}_j \equiv (\alpha_1p_1,..,
\alpha_np_n;\beta_1q_1,..,\beta_{j-1}q_{j-1},
 \beta_{j+1}q_{j+1},..,\beta_nq_n,\beta_j;\mu_1k_1,..,\mu_mk_m)$\\
and
\FL
\bey\label{compT}
&&{\cal T}^{(l)\L\L_0 }_{2n\,m}(\alpha_1p_1,..,
\alpha_np_n;\beta_1q_1,..,
\beta_nq_n;\mu_1k_1,..,\mu_mk_m) =\nonumber\\
&&(-1)^{n+1}\int_p K_\L(p)\Big[
S^{\Lambda\L_0}_{\beta\,\alpha}(p+P)L^{(l-1)\L\L_0}_{2n+2\:m}(\alpha_1p_1,..,\alpha_np_n,
\alpha\, -\!p;\beta_1q_1,..,\beta_nq_n,\beta;\mu_1k_1,..,\mu_mk_m)\nonumber\\
&&-L^{(l-1)\L\L_0}_{2n+2\:m}(\alpha_1p_1,..,\alpha_np_n,
\alpha\; p\!+\!P;\beta_1q_1,..,\beta_nq_n,\beta;\mu_1k_1,..,\mu_mk_m)
S^{\Lambda\L_0}_{\beta\alpha}(-p-P)\Big] 
\eny
with
$P=-\left(\sum_{j=1}^np_j+\sum_{j=1}^nq_j+\sum_{r=1}^mk_r\right)$.

Eq.\,(\ref{compO}) holds for $n>0$ and $l>0$. For $n=0$ the second
term
in the r.h.s. is absent.
For $l=0$ the r.h.s. of eq.\,(\ref{compO}), which now includes for  
$m=0$ and $n=1$ the
 term $ [-S^{-1}_{\alpha_1\,\beta_1}(-p_1)
+S^{-1}_{\alpha_1\,\beta_1}(q_1)]$ , is 
 zero due to the gauge invariance of the classical action. 

A little digression on the notation is in order: because of
translation invariance,
the $C^{\infty}$ functions $L_{2n\,m}^{\L\L_0}$ depend, for $\L > 0$, 
only on $2n + m - 1$ momenta.  In the case $n > 0$
we consider by convention $L_{2n\,m}^{\L\L_0}$ dependent on $n-1$
 momenta $q$ of the fermionic fields $\psi (q)$; therefore
eq.\,(\ref{compO}) (as well as other similar in the following)
appears to be asymmetric in the fermionic variables.
For $n = 0$ the functions 
$L_{0\,m}^{\L\L_0}$ depend on $m - 1$ bosonic momenta and
eq.\,(\ref{compO}) changes accordingly.

As a consequence of charge conservation the r.h.s. of
eq.\,(\ref{compO})
is equal to zero for $P = 0$ and thus the renormalization conditions
for ${\cal O}_{\L_R\L_0 }$ which do not include some derivative are vanishing.
We are then interested in
considering derivatives of order
$z > 0$ with respect to momenta.

Let us discuss briefly the various relevant sectors of eq.\,(\ref{compO}).

i) $m = 1,\ n = 0,\ z = 1,2,3$\,:\\
we can analyze all these cases by considering the Mc Laurin expansion
up to the third order in the momenta. The first addendum on the
r.h.s. of eq.\,(\ref{compO}) yields:\\
$-{1\over e}k_\mu L_{\mu\nu}^{(l)\L_R\L_0}(k)|_{rel}=
-{1\over e}k_\mu\left[z^{(l)}_3(k^2\delta_{\mu\nu}-
k_\mu k_\nu)+(\xi_1^{(l)}+\xi_2^{(l)}k^2)\delta_{\mu\nu}\right]
= -{1\over e}k_\nu(\xi^{(l)}_1 + \xi^{(l)}_2 k^2) $.\\
$O(4)$ symmetry and smoothness 
imply that ${\cal T}_{\nu}^{(l)\L_R\L_0}(k)|_{rel}$ has the same tensorial
structure as the r.h.s. of the previous equation, so that
by a suitable  choice of the renormalization constants $\xi_1^{(l)}$
and $\xi_2^{(l)}$
one imposes ${\cal O}_{\nu}^{(l)\L_R\L_0}(k)|_{rel} = 0$ (at one loop
see eq.\,(\ref{renph})).

ii) $m = 2,\ n = 0$\,:\\
invariance under charge conjugation (Furry theorem) leads to
$
L^{(l)\L_R\L_0}_{\mu \rho \sigma}( k_1, k_2)=0
$
and 
${\cal T}_{\rho \sigma}^{(l)\L_R\L_0}(k_1,k_2) = 0$: the r.h.s. of
eq. (\ref{compO})
vanishes identically.

iii) $m = 3,\ n = 0,\ z=1$\,:\\
$
{-{1\over e}(k_1+k_2+k_3)_\mu L^{(l)\L_R\L_0}_{\rho\sigma\tau\mu}}|_{rel}=
-{1\over e}(k_1+k_2+k_3)_\mu
\delta_{(\mu\rho} \delta_{\sigma\tau)}\xi^{(l)}_4
$\\
${\cal T}_{\rho \sigma \tau}^{(l)\L_R\L_0}(k_1,k_2,k_3)|_{rel}$\  
has the same structure
as the r.h.s. of the previous equation; indeed by
applying ${\partial \over {\partial k_{1_{\nu}}}}$ for $k_1=k_2=k_3=0$
to the r.h.s. of eq.\,(\ref{compT}) for $n=0$ and $ m=3$,
one notices that the only non vanishing contributions arise  when the 
derivative act on the $P=-(k_1+k_2+k_3)$ variable. Namely to the
first
order ${\cal T}_{\rho \sigma \tau}^{(l)\L_R \L_0}$ depends on the momenta
only through $P$, then the bosonic and $O(4)$  symmetries lead
easily to the conclusion. Therefore with a 
suitable choice of the renormalization coefficient $\xi_{4}^{(l)}$
(at one loop see eq.\,(\ref{RCL}))
it is possible to set ${\cal O}_{\rho \sigma \tau}^{\L_R\L_0} |_{rel} = 0$.

iv) $ m=0,\ 2n=2, \ z=1$: \\
for $l>0$ the first two addenda on the r.h.s. of eq.\,(\ref{compO})
give\\
${\left\{-{1\over e}(p+q)_\mu L_{\alpha\beta\mu}^{(l)\L_R\L_0}(p;-p-q)
+L^{(l)\L_R\L_0}_{\alpha\beta}(-q)
-L^{(l)\L_R\L_0}_{\alpha\beta}(p)
\right\}} |_{rel}=
 -i(z_1^{(l)}-z^{(l)}_2)(\ds{p}+\ds{q})_{\alpha\beta} $\\
where $z_1^{(l)}$ and $z^{(l)}_2$ are the renormalization constants
already introduced for $l=1$ in sect.\,II.
Using charge conjugation invariance it is easy to show that
$ {\cal T}_{\alpha \beta}^{(l)\L_R\L_0}(p,q)$ has the same structure 
so that 
$ {\cal O}_{\alpha \beta}^{(l)\L_R\L_0} |_{rel} = 0$ can be satisfied 
with a suitable choice of 
$z_1^{(l)}-z^{(l)}_2$ (see eq.\,(\ref{Wtwo}).

 Observe that
the effective Ward identities determine the renormalization conditions
up to the two arbitrary constants $z_1$ and $z_3$.
\vskip 0.4cm
After this discussion a formal proof of ultraviolet and infrared finiteness
of QED can be made along these lines:
 one wants to prove a series of suitable bounds concerning the
 ultraviolet and infrared behavior of $L_{2n\ m}^{(l)\L\L_0} $,
 in an inductive scheme in the
loop index $l$.
Because of our choice of the cut-off function $K$ we can use the results
of \cite{KKQED2,KK}.  The infrared finiteness of a 
theory with massless fermions and 
photons has been proven in \cite{KKQED2}, independently of any Ward
identity, provided the renormalization conditions 
\bea\label{RC}
L_{\alpha\beta}^{(l)\, 0\L_0} (0) =
L_{\mu \nu }^{(l)\, 0\L_0} ( 0)=0
\ena
 are imposed.
  We shall see that, for arbitrary values of $z_1$ and $z_3$,
 eqs.\,(\ref{RC}) are satisfied.

 The ultraviolet part of the proof is standard, ultraviolet
 finiteness following, for \hbox {$\L > 0$}, from arbitrary renormalization
 conditions at $\L = \L_R$, not necessarily compatible with the Ward
 identities.
 From the U-V bounds for $L_{2n\ m}^{(l)\L\L_0} $ \cite{KKS} one obtains as
a consequence the suitable power-counting  bounds for the irrelevant
components of ${\cal O}^{(l)}_{ \L_0 } $ of eq.\,(\ref{defO}) that,
together
with ${\cal O}^{(l)}_{ \L_R\L_0}|_{rel}=0$, lead to
$lim_{\L_0 \to \infty}{\cal O}^{(l)}_{ \L \L_0}[\Phi]=0 $.
Therefore for $\L>0$ and loop $l$ we can perform the
limit $\L_0 \to \infty$ in eq.\,(\ref{idO}), so that the effective
Ward identities are satisfied.

We have to discuss now at loop $l$
the $\L \to 0$ limit.
From the infrared bounds on the vertices \cite{KKQED2,KK}, which are valid by 
induction hypothesis  at loop order \hbox{$l' \leq l - 1$}, and using
eq.\,(\ref{compT}) one has
$lim_{\L\to 0}\;\partial_{k_{\mu}} {\cal T}_{\nu}^{(l) \L \L_0= \infty} (k)
|_{k=0}=0 $.
 Using eq.\,(\ref{compO}) with \hbox{$m=1,\ n=0,\ z=1\ $} one gets
$lim_{\L\rightarrow 0}\;L_{\mu \nu}^{ (l) \L\L_0= \infty} (0) = 0$. 
Moreover the renormalization conditions on 
$L_{\alpha\beta}^{ (l)\L_R\L_0} (0)$ has not been involved in the
 effective Ward identity (\ref{compO}), so that we can choose
 \hbox {$L_{\alpha\beta}^{ (l)\L_R\L_0} (0)=0$}; from  rigid axial
 symmetry it follows that $L_{\alpha\beta}^{ (l)\L\L_0} (0)=0$ for any $\L$.
Therefore eqs.\,(\ref{RC})  are satisfied and one can prove all the I-R bounds
at loop $l$
 as required by the induction scheme, so that finiteness together with all the 
 effective Ward identities are proved; moreover for non-exceptional momenta 
 the ${\cal T}^{(l)}$ functions of eq.\,(\ref{compT}) go to zero as 
$\L \rightarrow 0$, and one recovers the usual Ward identities.
 
Notice that this renormalization procedure is not 
sufficient in massless scalar  QED to ensure
infrared finiteness, since there is no chiral symmetry protecting
the scalar from getting a mass. Imposing renormalization conditions
compatible with the effective Ward identities, the theory is
ultraviolet finite and satisfies the effective Ward identities for
$\L>0$ and $\L_0 \to \infty$, but infrared finiteness requires  
that the renormalization conditions on $L_{\phi^2}^{(l)\L_R\L_0}(0) $
must be chosen in such a way that 
$ \lim_{\L \to 0} L_{\phi^2}^{(l)\L\L_0}(0) =0 $. 
This fine tuning is typical for theories with
massless scalars.

\subsection{Gauge invariance of composite operators}
 In this subsection we will shortly discuss the gauge invariance of 
 composite operators. One associates to the gauge variation of
 a local composite operator $J(x)$ the insertion:
\bey\label{defOJ}
{\cal O}_{\L\L_0}^{J}[x;\Phi;\omega] =&&
i\int dx'\,\omega(x')\,{\cal O}_{\L\L_0}^{J}[x',x;\Phi]  \nonumber\\=&&
e^{L_{\L\L_0}[\Phi]}\, e^{\Delta_{\L\L_0}\left(\frac
 {\delta}{\delta\Phi}\right)}\left[(R\Phi +T)K_{\L_0}\frac{\delta}{\delta\Phi}-
\Phi D^{-1}R\Phi\right]\left(J_{\L_0}[x;\Phi]\,e^{-L_{\L_0}[\Phi]}\right)
\eny
$J_{\L_0}[x;\Phi]$ being the bare composite operator, boundary
 condition
of eq.\,(\ref{fluxO}).
By commuting the argument of the square brackets with the exponential of
the Laplacian we easily get the following expression in terms of the functional
$J_{\L\L_0}[x;\Phi]$, solution of eq.\,(\ref{fluxO}):
\FL\bey\label{idOJ}
&&{\cal O}_{\L\L_0}^{J}[x;\Phi;\omega]
 = 
TK_{\L_0}J^{\,'}_{\L\L_0}[x;\Phi]
-\Phi D^{-1}RD_{\L\L_0}J^{\,'}_{\L\L_0}[x;\Phi]
+\Phi R^TK_{\L}J^{\,'}_{\L\L_0}[x;\Phi]\nonumber\\&&
-J^{\,'}_{\L\L_0}[x;\Phi]\left(D_{\L\L_0}R^TK_{\L}+
K_\L RD_{\L\L_0}\right)L_{\L\L_0}^{\,'}[\Phi] 
+ {\cal T}_{\L\L_0}^J[x;\Phi;\omega]
+{\cal O}_{\L\L_0}[\Phi;\omega]J_{\L\L_0}[x;\Phi]
\eny
 where
 \bea\label{defTJ}
 {\cal T}_{\L\L_0}^J[x;\Phi;\omega]=trK_\L RD_{\L\L_0} J^{\,''}_{\L\L_0} [x;\Phi]
 \ena
 The primes mean as usual differentiation with respect to
 $ \Phi $; $ {\cal O}_{\L\L_0}$ is the functional corresponding to the
evanescent operator discussed
in the previous subsection.

To show that a composite operator $J$ is gauge invariant, one has to prove
that it is possible to impose the renormalization conditions on $J$ in such a
way that $ {\cal O}^J_{\L\L_0}$ is evanescent. In particular
if $J_{\L_0}$ is gauge invariant at classical level one checks easily
from  eq.\,(\ref{defOJ}) that ${\cal O}^J_{\L\L_0}$ is evanescent at tree level.
We stress that by gauge invariant operator we mean an operator
satisfying
the effective Ward identities; we do not address the question of its
independence from the gauge fixing parameter.

Taking the functional derivative with respect to the fields and the
 gauge
parameter $\omega$, in a suitable neighbourhood of the origin of the
momentum space, eqs.\,(\ref{idOJ}) and (\ref{defTJ}) yield for $l>0$ (modulo
evanescent terms):
\bey \label{compOJ}
 {\cal O}&&^{(l)\L\L_0 }_{J\;2n\,m}(\alpha_1p_1,\ldots ,
 \alpha_np_n;\beta_1q_1,.., 
 \beta_nq_n;\mu_1k_1,..,\mu_mk_m;Q)=\nonumber \\&&
 -{1\over e}P_{\mu}J^{(l)\L\L_0}_{2n\,m+1}
(\alpha_1p_1,.., \alpha_np_n;\beta_1q_1,
 .., \beta_{n-1}q_{n-1},\beta_nq_n;\mu_1k_1,..,\mu_mk_m,\mu P)\\
&&-\sum_{j=1}^{n}\Big[
J^{(l)\L\L_0}_{2n\,m}({\cal P}_j)
-J^{(l)\L\L_0}_{2n\,m}({\cal Q}_j)\Big]+{\cal T}^{(l)\L\L_0}_{J\,2n\,m}(\alpha_1p_1,..,\alpha_np_n;\beta_1q_1,
 ..,\beta_nq_n;\mu_1k_1,..,\mu_mk_m;Q)\nonumber
 \eny
 where the multi-indices ${\cal P}_j$ and ${\cal Q}_j$ are defined
 by\\
${\cal P}_j \equiv (\alpha_1 p_1,.., \alpha_{j-1} p_{j-1},\alpha_j\,P\!+\!p_j,
\alpha_{j+1}p_{j+1},..,\alpha_np_n;\beta_1q_1,
 ..,\beta_nq_n;\mu_1k_1,..,\mu_mk_m)\\
{\cal Q}_j \equiv (\alpha_1 p_1,..,\alpha_np_n;\beta_1q_1,..,
\beta_{j-1}q_{j-1},
\beta_j\,P\!+\!q_j,\beta_{j+1}q_{j+1},\beta_nq_n;\mu_1k_1,..,\mu_mk_m)$\\
and
 \FL
 \bey\label{compTJ}
 {\cal T}&&^{(l)\L\L_0}_{J\,2n\,m}(\alpha_1p_1,..,
 \alpha_np_n;\beta_1q_1,..,
 \beta_nq_n;\mu_1k_1,..,\mu_mk_m;Q) =(-1)^{n+1}\int_p \bigg\{K_\L(p)\times\nonumber\\
 &&\times \Big[ S^{\L\L_0}_{\beta\,\alpha}(p+P) 
J_{2n+2\,m}^{(l-1)\L\L_0}(\alpha_1p_1,..,
\alpha_np_n,\alpha\,-\!p;\beta_1q_1,..,\beta_nq_n,
\beta\,P\!+\!p;\mu_1k_1,..,\mu_mk_m)\\
 &&-J_{2n+2\,m}^{(l-1)\L\L_0}(\alpha_1p_1,..,\alpha_np_n,
 \alpha\, p\!+\!P;\beta_1q_1,..,\beta_nq_n,\beta\, -\!p;\mu_1k_1,..,\mu_mk_m)
 S^{\L\L_0}_{\beta\alpha}(-p-P)\Big]\bigg\}\nonumber 
 \eny
with $P=-(\sum_{j=1}^n p_j + \sum_{j=1}^n q_j +\sum_{r=1}^m k_r +Q)$.\\
 In eqs.\,(\ref{compOJ}) and (\ref{compTJ}) $Q$ is the momentum conjugated to the 
$x$ variable in 
eq.\,(\ref{idOJ}); notice that the arguments of the $J$ functions
 refer to the fields variables, their sum is therefore always equal
to $-Q$. 

We shall discuss the cases  
 $J(x) = i\bar{\psi} \gamma^{\tau} \gamma_5 \psi (x) 
 \equiv J^{\tau}_{ 5}(x)$
 and $J(x) = F\tilde{F}(x)$, the former renormalized as an operator of
 dimension $3$, the latter as an operator of dimension $4$.

For $J_{5}^{\tau}$ the relevant projections of 
${\cal O}^{J}$ are:\\
$n=0,\ m=1,\
 z=0,1,2;\ \ n=0,\ m=2,\ z=0,1;\ \ n=0,\ m=3,\ z=0;$ \\
$n=1,\ m=0,\ z=0$.\\
The rigid symmetries constrain $J^{\tau}_{5 \L_0}$ to be a linear
  combination of 
 $\bar{\psi} \gamma^{\tau} \gamma_5 \psi$ and 
 $\epsilon_{\tau \nu\rho \sigma}F^{\nu \rho} A^{\sigma}$.

\noindent The $z=0$ projections are identically satisfied because of charge
 conservation.

\noindent For $m=1,\ n=0,\ z=1$ all the addenda of eq.\,(\ref{compOJ})
vanish, 
since no
pseudotensor with three indices exists.

\noindent For $m=1, \ n=0, \ z=2$ there is one possible structure, the pseudotensor
 $\epsilon_{\tau\rho\mu\nu}$. Thus the completely antisymmetric
 tensor
 $\frac{\partial}{\partial k_{1\rho}}J_{5\mu\nu}^
{\tau(l)\L_R\L_0}(k_1,k_2)|_{k_1=k_2=0}$
 apart from some trivial numerical constant is to be chosen equal to
 $ \frac{\partial}{\partial k_{1\rho}}\frac{\partial}{\partial
 Q_{\mu}}{\cal T}_{5\nu}^{\tau (l)\L_R\L_0}(k_1;Q)|_{k_1=Q=0}$
 which does not vanish as shown in sect.\,II.

\noindent The projections $m=2,\ n=0$ involve
 $ (q_1+q_2+Q)_\mu J^{\tau (l)\L_R\L_0}_{5\nu\rho\mu}(q_1,q_2,-q_1-q_2-Q) $
 and ${\cal T}_{5\nu \rho}^{\tau (l)\L_R\L_0}$ and both, because of 
charge conjugation 
 invariance, are identically zero.  

\noindent The identity for $m=3, \ n=0, \ z=0$
 is satisfied, as well as for $2n=2, \ m=0, \ z=0$.

 The conclusion of our analysis is that 
$J_{5\L\L_0}^{\tau}[x;\Phi]$
 is unambiguosly
 determined 
up to a multiplicative constant which 
is fixed by the renormalization condition 
 \bea\label{RCJ5}
J_{5 \alpha \beta}^{(l)\tau \L_R\L_0}(0) = i\sigma^{(l)} ( \gamma^{\tau} 
  \gamma^{5} )_{\alpha \beta}
\ena
the mixing between the two 
possible bare
operators being determined loop by loop by the requirement of gauge
invariance.

One can introduce at scale $ \L_R $ for an operator of dimension $d$ 
 an analogue of the
 Zimmermann normal product, which we call
 $N_{d}^{\L_R\, c}[J]$ (c meaning connected rather than proper),
by choosing for $l > 0$ vanishing renormalization conditions at $\L=\L_R $ 
 and zero momentum \cite{Zimm,Becchi2,KKop}.
Notice that for $J_5^\tau$ the renormalization recipe  
$N_3^{\L_R\,c}[J^\tau_5]$ is not gauge invariant, as we saw explicitly
in eq.\,(\ref{div}).
The composite operator so defined is infrared finite, namely the limit
 $\L \rightarrow 0$ exists for non-exceptional momenta; in fact, as already
 stated, the positive dimension Green functions $J_{5\mu}^{\tau \L\L_0}(0)$,
 $J_{5\mu \nu}^{\tau \L\L_0}(0,0)$ and 
$\partial_{\rho} J_{5\mu}^{\tau \L\L_0}(0)$
  are zero due to the rigid   symmetries of theory 
 (the condition analogue of eq.\,(\ref{RC}) is then satisfied);
 as a consequence ${\cal T}^{\tau}_{5\L\L_0}[x;\Phi] \rightarrow 0$
 as $\L \rightarrow 0$ and in this limit we recover the usual Ward identity
 for the composite operators.  

As a last remark we note that due to linearity of eq.\,(\ref{idOJ}) 
${\partial\over{\partial x_{\tau}}} J_{5}^{\tau\L\L_0}[x,\Phi]$ 
fulfils the effective Ward identity too, in the limit $\L_0 \to \infty $. 

 Similar considerations can be
 repeated for the renormalization of $F \tilde{F}$ (which mixes with
 $\partial_{\mu}( \bar{\psi} \gamma_{\mu} \gamma_{5} \psi )$ )
 but now the result is simpler: indeed $N_{4}^{\L_R\,c}[F \tilde{F} ]$
 yields a gauge invariant renormalization.

\subsection{Effective axial Ward identity}

 Let us consider the insertion ${\cal O}^{\,5}_{\L\L_0}[x,\Phi]$ defined
by eq.\,(\ref{defO})
with  $T = 0$ and $R = R_{5}$ parameter of a local axial 
 transformation.  At tree level,
by the naive Noether construction, it is equal to
$ -i\partial_{\mu} J_{5\,\L\L_0}^{\mu}[x,\Phi]$; at quantum
level 
$$i{\cal O}^{\,5}_{\L\L_0}[x;\Phi]
 -\partial_{\mu} J_{5\,\L\L_0}^{\mu}[x;\Phi]\equiv
i{\cal A}_{\L\L_0}[x;\Phi]
$$ 
represents the anomaly term.
Proceeding as in (\ref{defO}) and (\ref{idO}) we get:
\FL
 \bey\label{idOJ5}
 i\int{dx}\,\omega_5(x)\,{\cal A}&&_{\L\L_0}[x;\Phi]=
-\int{dx}\,\omega_5(x)\,\partial_\mu J_{5\L\L_0}^\mu[x;\Phi] 
-\Phi D^{-1}R_5\Phi 
+\Phi D^{-1}R_5D_{\L\L_0}L^{'}_{\L\L_0}[\Phi]\nonumber\\&&
 -\Phi R_5^TK_\L L^{\,'}_{\L\L_0}[\Phi]
-{L'}_{\L\L_0}[\Phi] D_{\L\L_0}R_5^TK_\L {L'}_{\L\L_0}[\Phi]
 -{\cal T}_{5\L\L_0}[\Phi;\omega_5]
\eny
 where
 \bea\label{defT5}
 {\cal T}_{5\L\L_0}[\Phi;\omega_5]=trK_\L R_5 D_{\L\L_0} L^{''}_{\L\L_0} [\Phi]
 \ena
 We stress that the composite operator ${\cal A}_{\L\L_0}[x;\Phi]$ 
satisfies the flow equation (\ref{fluxO}), indeed both 
${\cal O}^{\,5}_{\L\L_0}[x;\Phi]$
 and $\partial_{\mu} J_{5\L\L_0}^{ \mu}[x;\Phi]$
solve this linear equation.

Let us show that, by a suitable choice of the 
renormalization constant $\sigma^{(l)}$ 
of eq.\,(\ref{RCJ5})  
 \bea
 \lim_{\L_0\rightarrow\infty}{\cal
 A}_{\L\L_0}[x;\Phi]=a\lim_{\L_0\to\infty}
 {\left[N_4^{\L_Rc}[F\tilde F]\right]}_{\L\L_0}[x;\Phi]
 \ena
  where $a$ is the coefficient of the anomaly, which is independent
from $\L$ and $\L_R$.

  The operator ${\cal A}_{\L\L_0}$ of dimension $4$ 
is defined by its renormalization
  conditions, which from eq.\,(\ref{idOJ5}) are determined by those
of $L_{\L_R\L_0}$ and $J^{\mu}_{5\L_R\L_0}$.
 
Consider eq.\,(\ref{idOJ5}) in a suitably small
neighbourhood of the
  origin of the momenta and for $\L = \L_R$ and $l>0$:
 \bey \label{compA}
 {\cal A}^{(l)\L_R\L_0 }_{2n\,m}(\alpha_1p_1,&&..,
 \alpha_np_n;\beta_1q_1,.., 
 \beta_nq_n;\mu_1k_1,..,\mu_mk_m)=\nonumber\\
=&&\sum_{j=1}^{n}(-1)^{n-j}
 \big[\gamma^5_{\alpha_j\,\alpha}L
 ^{(l)\L_R\L_0}_{2n\,m}({\cal P}_{j\alpha})
 +L^{(l)\L_R\L_0}_{2n\,m}({\cal Q}_{j\beta})
\gamma^5_{\beta\beta_j}\big]\\&&
 -\big[{\cal T}^{(l)\L_R\L_0}_{5\,2n\,m}
+P_\tau J^{(l)\tau\L_R\L_0}_{2n\,m}\big](\alpha_1p_1,..,
\alpha_np_n;\beta_1q_1,.., 
\beta_nq_n;\mu_1k_1,..,\mu_mk_m)\nonumber 
 \eny
 where ${\cal P}_{j\alpha}$ and $ {\cal Q}_{j\beta}$ are defined by\\
${\cal P}_{j\alpha}\equiv (\alpha_1 p_1,.., \alpha_{j-1}p_{j-1},
\alpha_{j+1}p_{j+1},..,
\alpha_np_n,\alpha\;P+p_j;\beta_1q_1,
 ..,\beta_{n-1}q_{n-1},\beta_n;\mu_1k_1,..,\mu_mk_m)\\
{\cal Q}_{j\beta}\equiv (\alpha_1 p_1,..,\alpha_np_n;\beta_1q_1,
..,\beta_{j-1}q_{j-1},\beta_{j+1}q_{j+1},..,\beta_nq_n,
\beta;\mu_1k_1,..,\mu_mk_m)$\\
and
 \FL
 \bey\label{compT5}
 {\cal T}&&^{(l)\L_R\L_0}_{5\,2n\,m}(\alpha_1p_1,..,\alpha_np_n;\beta_1q_1,
 ..,\beta_nq_n;\mu_1k_1,..,\mu_mk_m) =
(-1)^{n+1}\int_p\bigg\{K_{\L_R}(p)\times\nonumber\\
 &&\times\Big[
  S^{\L_R\L_0}_{\beta\,\alpha}(p+P)\gamma^5_{\alpha\alpha'}
 L_{2n+2\,m}^{(l-1)\L_R\L_0}(\alpha_1p_1,..,\alpha_np_n,\alpha'\,-\!p;
 \beta_1q_1,..,\beta_nq_n,\beta;\mu_1k_1,..,\mu_mk_m)\\
 &&+L_{2n+2\,m}^{(l-1)\L_R\L_0}(\alpha_1p_1,..,\alpha_np_n,
 \alpha \,p\!+\!P;\beta_1q_1,..,\beta_nq_n,\beta';\mu_1k_1,..,\mu_mk_m)
 \gamma^5_{\beta'\beta}S^{\L_R\L_0}_{\beta\alpha}(-p-P)\Big]\bigg\}\nonumber 
 \eny 
 with $ P=-\left(\sum_{j=1}^n p_j+\sum_{j=1}^n q_j+\sum_{r=1}^m k_r\right)$.
As regards the momentum dependence in eq.\,(\ref{compA}) 
considerations similar to those after eq.\,(\ref{compO}) hold.
When we consider the relevant projections in eq.\,(\ref{compA}) 
we find, using
  the rigid symmetries of $L_{\L\L_0}$ and of $J_{5\L\L_0}^{\tau}$,
  that all the renormalization conditions of ${\cal A}_{\L_R\L_0}$ are zero
  identically (i.e. independently of the renormalization conditions
  on $J_{5}^{\tau}$ ) but the following two:

i) $2n =2,\ m=0, \ z=1 $:\\ the divergence of the current on the r.h.s. gives 
 $i(\ds{p} + \ds{q}) \gamma_{5} \sigma^{(l)}$, where $\sigma^{(l)}$
  is the (arbitrary for the moment) renormalization constant in 
eq.\,(\ref{RCJ5}).
  The first addendum in the r.h.s. of eq.\,(\ref{compA}) gives an
  analogous term
$  \left[\gamma^5_{\alpha\alpha'}L^{(l)\L_R\L_0}_{\alpha'\beta}(-q)+
  L^{(l)\L_R\L_0}_{\alpha\beta'}(p)\gamma^5_{\beta'\beta}\right]|_{rel}=-i
  z^{(l)}_2\left[(\ds{p} + \ds{q}) \gamma_{5}\right]_{\alpha\beta}$.
Using  charge conjugation invariance of $L_{\L\L_0}$ one checks that 
$ {\cal T}_{5\alpha \beta }^{(l)\L_R\L_0}(p,q)|_{rel} = 
it^{(l)}\left[(\ds{p} + \ds{q}) \gamma_{5}\right]_{\alpha\beta}$,
 where $t^{(l)}$ depends on $L^{(l')}_{\L\L_0}$ at loop $l' < l$.
Fixing loop by loop   
 $\sigma^{(l)}$ in terms of $z_{2}^{(l)}$
we can set to zero the corresponding renormalization condition
 on ${\cal A}^{(l)}_{\L_R\L_0}$.

ii) $m=2,\ n=0,\ z=2$:\\
the divergence term
$
{\partial \over{\partial k_{1\rho}}}{\partial \over{\partial
k_{2\sigma}}}
(k_1+k_2)_\tau J_{5\mu\nu}^{\tau(l)\L_R\L_0}(k_1,k_2)|_{k_1=k_2=0}
=2{\partial \over{\partial k_{1\rho}}}J_{5\mu\nu}^{\sigma(l)\L_R\L_0}
(k_1,k_2)|_{k_1=k_2=0} $
 as previously discussed is not vanishing because of gauge invariance,
but also  the term
${\partial^2 \over {\partial k_{1\rho}  \partial k_{2\sigma}}}
{\cal T}_{\mu \nu}^{(1)\L_R\L_0}(k_1,k_2)|_{k_1=k_2=0}$
 is not vanishing, both the quantities for $O(4)$ symmetry and parity
 being proportional to $\epsilon_{\rho \sigma \mu \nu}$ . At one loop
 the two terms, both proportional to the same integral, were computed
 in sect.\,II; 
 they sum up to give the well-known coefficient of the anomaly.
 By direct inspection on the renormalization conditions
 of ${\cal A}$ and comparison with those  of 
$N_4^{\L_Rc}[F \tilde{F} ]$
 we can state that for $\L_0 \rightarrow \infty$
 the two functional are proportional. 
Since ${\cal A}_{\L\L_0}$ and $\left[N_4^{\L_Rc}[F
\tilde{F}]\right]_{\L\L_0}$
satisfy eq.\,(\ref{fluxO}), $a$ is independent of $\L$; for
dimensional
reasons it is also $\L_R$-independent.

\section*{Conclusions}
Polchinski has shown that the Wilsonian renormalizaton group can be
applied to get a rigorous proof of renormalizability, which avoids the
subleties involved in the BPHZ technique.

In this paper we show that this method provides a natural
generalization of the hard-soft renormalization program suggested in
\cite{LM}, in which the renormalization scale $\L_R$ substitutes the
momentum subtraction scale $\mu$ as the only dimensional parameter
introduced in the renormalization of massless theories.
To prove that physical quantities are independent from $\L_R$, it is
necessary to find the Gell-Mann and Low renormalization
group equation in the HS schemes, discussed in 
\cite{LM,HL}. In this sense the coupling constants in HS
schemes are as `physical' as in the non-zero momentum subtraction scheme.
Our modification of the HS scheme introduced in \cite{LM} has 
the advantage of requiring renormalization
conditions only at $\L_R$, while in \cite{LM} there are also 
renormalization conditions at $\L=0$; furthermore, renormalizability
is investigated using the method of the flow equation instead of BPHZ.

As in the usual BPHZ treatment, the validity of Ward identities is
established associating to them a composite operator which, if the
Ward identities are not anomalous, can be proven to be evanescent;
otherwise it is a local operator, the anomaly. The HS scheme has the
advantage that this can be done computing simply zero-momentum graphs.

As an example of the application of this method, we consider the case
of massless QED.
Using the flow equation and some results of
\cite{KKQED2}, we prove the validity of the effective Ward 
identities for massless QED at quantum level; then we show that, for
any HS scheme  compatible with the
effective Ward identities and the rigid symmetries, the theory is
infrared finite.
We define in the same formalism the gauge-invariant axial current
operator and its anomaly.

In the case of Yang-Mills the effective Ward identities have been proven in
\cite{Becchi2} in a HS scheme; it would be interesting to prove 
infrared finiteness in the same scheme.

\subsection*
{\centerline{\it Note Added in Proof}}

To prove the evanescence of ${\cal O}_{\L \L_0}^J$ in eq.(\ref{defOJ}) , one
cannot use directly its flow equation (\ref{fluxO}), since it is a
disconnected insertion; however its connected part 
${\cal O}_{\L \L_0}^J - {\cal O}_{\L \L_0} J_{\L \L_0}$ satisfies a
modified flow equation, which differs from  eq.(\ref{fluxO}) by an
inhomogeneous term 
${\cal O}_{\L \L_0}' \partial_{\L} D_{\L \L_0}^T J_{\L \L_0}'$.
Using this modified flow equation and the fact that ${\cal O}_{\L
\L_0}$ is evanescent (see Subsection III.A) the proof of evanescence of
${\cal O}_{\L \L_0}^J - {\cal O}_{\L \L_0} J_{\L \L_0}$ , and hence of
${\cal O}_{\L \L_0}^J$, is essentially the same as in the standard one
for a connected operator insertion satisfying eq.(\ref{fluxO}).

\end{document}